\documentclass[aps,prl,reprint,groupedaddress]{revtex4-1}

\newcommand{\pe}{\mbox{Pe}}

\newcommand{\dt}[1]{\frac{\partial #1}{\partial t}}

\usepackage{graphicx}
\usepackage{amssymb,amsfonts,amsmath,times,color}
\usepackage[english]{babel}

\begin{document}
\title{Localization and dynamics of sulfur-oxidizing microbes in natural sediment}
\author{Alexander Petroff$^{1*}$, Frank Tejera$^1$ \& Albert Libchaber$^1$}
\email{apetroff@rockefeller.edu} \affiliation{$^{1}$Laboratory of
  Experimental Condensed Matter Physics, The Rockefeller
  University. New York City, NY, USA}

\begin{abstract}

  Organic material in anoxic sediment represents a globally significant carbon reservoir that acts to stabilize Earth's atmospheric composition. The dynamics by which microbes organize to consume this material remain poorly understood. Here we observe the collective dynamics of a microbial community, collected from a salt marsh, as it comes to steady state in a two-dimensional ecosystem, covered by flowing water and under constant illumination. Microbes form a very thin front at the oxic-anoxic interface that moves towards the surface with constant velocity and comes to rest at a fixed depth. Fronts are stable to all perturbations while in the sediment, but develop bioconvective plumes in water. We observe the transient formation of parallel fronts. We model these dynamics to understand how they arise from the coupling between metabolism, aerotaxis, and diffusion. These results identify the typical timescale for the oxygen flux and penetration depth to reach steady state.

\end{abstract}

\maketitle



When organic material is buried in sediment~\cite{hayes2006carbon},
its decay by microbes is slowed by the limited diffusive flux of
oxygen from the surface~\cite{fenchel2012bacterial}, thus sequestering
a large reservoir of fixed
carbon~\cite{lal2008carbon,bridgham2006carbon}.
The burial of organic material contributed to the rise of oxygen in
the ancient atmosphere~\cite{catling2005earth,dismukes2001origin} and
its decay may further destabilize the modern
climate~\cite{koven2011permafrost}.
Predicting the quantity carbon sequestered and respired requires an
understanding of how complex microbial communities organize and move in
nutrient gradients.
Although the collective dynamics of simple microbial communities are
well
studied~\cite{wu2000particle,zhang2010collective,platt1961bioconvection,douarche2009coli,petroff2014hydrodynamics,petroff2015fast},
the collective dynamics of complex microbial communities---typical of
natural sediment--- remain poorly understood.
Here we observe the dynamics by which a microbial community, collected
from a salt-marsh microbial mat and brought into the lab, organizes to
fix the oxygen penetration depth.
At this depth, sulfur-oxidizing bacteria consume oxygen and H$_2$S
(sulfide)~\cite{robertson1992colorless}, the waste product of
sulfur-reducing bacteria~\cite{widdel1992dissimilatory}. As these
reducing and oxidizing bacteria exchange sulfur compounds, they
maintain a sulfur
cycle~\cite{fenchel2012bacterial,gregor1988chemical}.
Periodic observations by Garcia-Pichel \textit{et al} have shown that
sulfur-oxidizing bacteria organize into a front that moves through the
sediment~\cite{garcia1994diel}.
We first describe a new experiment that allows one to continuously
observe these dynamics in a two-dimensional chamber.
We find that microbes self-organize into a thin stable front that
moves with constant velocity to a steady-state depth.
We then present a model describing how these dynamics arise
from the coupling between aerotaxis, metabolism, and diffusion.
These results allow us to identify the typical timescale required for
the oxygen penetration depth to come to steady-state in natural
sediment.

\begin{figure*}
\centering \includegraphics[width=1\textwidth]{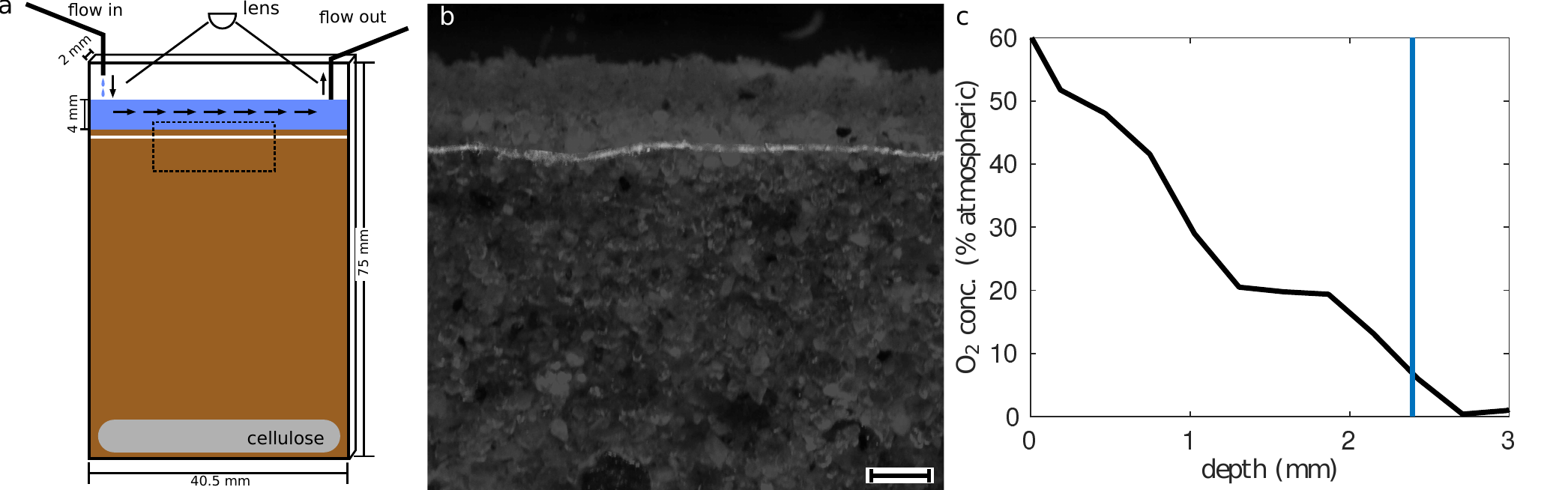}
\caption{Microbes in natural sediment organize into metabolically
  distinct strata. (a) The dynamics of sulfur-oxidizing bacteria are
  observed in a Hele-Shaw cell with a flow of fresh media at the
  sediment surface. (b) A white band of sulfur-oxidizing bacteria
  forms. The front, here positioned at its steady-state depth, extends
  the entire width of the chamber. The chamber is imaged at 15 minute
  intervals. Scale bar is 1 mm. (c) Measurement of the oxygen
  concentration around the front (blue line) at the moment of its
  formation show it positioned near the oxic-anoxic
  interface.}\label{front_1}
\end{figure*}

We begin by collecting a cyanobacterial mat from Little Sippewissett
Marsh ($41.575268^{\circ}\ $N $70.638406^{\circ}\ $W) near Woods Hole
Massachusetts~\cite{nicholson1987structure,buckley2008vertical}.
Bringing this material into the lab, we observe the typical metabolic
stratification. A green stratum of photosynthetic microbes lays above
purple, white, and dark bands of purple sulfur, sulfur-oxidizing, and
sulfur-reducing bacteria,
respectively~\cite{pierson1987pigments,fenchel2012bacterial}.
Details regarding the collection and maintenance of this mat in the lab
are described in reference~\cite{petroff2017}.

To continuously observe the motion of the sulfur-oxidizing microbes in
a constant environment, we developed a two-dimensional microbial
ecosystem in a Hele-Shaw chamber~\cite{saffman1958penetration}.
Figure 1(a) shows a schematic of this experiment.
We place sediment between two $40.5\ \mbox{mm} \times 75\ \mbox{mm}$
clear acrylic walls separated by a 2 mm ($\sim10\ $sand grains) rubber
spacer, held together by bolts at the edges of the chamber.
We continuously flow fresh saltwater media over the surface of the
sediment at a flow rate of $Q=0.44$ cm$^3/$min (oxygen Peclet number
$\pe\sim 5000$), which maintains a constant chemical environment at
the surface.
The oxygen concentration in the fresh media is in equilibrium with the
atmosphere and contains $20\ $mM SO$_4^{2-}$, typical of natural salt
marshes~\cite{goldman1978steady,petroff2014hydrodynamics}. These
oxidants maintain populations of aerobic, sulfur-oxidizing, and
sulfur-reducing bacteria in the sediment.
To mimic the natural environment, we focus a light sheet (produced
from a high powered Thorlabs MCWHL5 led) onto the top surface of the
sediment while keeping the edges of the chamber in darkness.

The way the chamber is loaded is critical to the experiment.
We take sediment from a microbial mat by making two parallel slices
into the sediment $5\ $mm apart with clean metal knives. We take the
sediment from between the slices by pressing the knives together and
lifting.
We lay this sediment on one of the acrylic walls surrounded on three
sides by the rubber spacer.
Extra sediment is removed to the height of the rubber spacer. We then
place the second acrylic wall onto the sediment.
Finally, we bolt the walls together to avoid leaks.
This compresses the sediment.
Although we are careful to preserve the vertical structure of the
sediment as we lay it on the chamber wall, the topmost several
millimeters of the mat mix. The thickness of the mixed layer varies
between experiments.
As a result, there is spurious oxygen in the top most layer of
sediment.
In these experiments, we observe the dynamics by which this oxygen
reservoir is consumed as the oxic-anoxic interface comes to steady
state.

5 to 8 hours after loading the chamber, we observe the formation of an
extraordinarily thin front of microbes (fig~1b), with a typical
thickness of only $120\pm60\ \mu$m. It extends the entire $40.5\ $mm
width of the chamber.
Due to the presence of sulfur globules~\cite{robertson1992colorless}
in the constituent bacteria, this band is clearly visible to the eye,
appearing as a thin white band.
This allows us to follow the dynamics of this front by photographing
its position at $15\ $minute intervals.

We observe a wide range of initial positions of the front (fig.~2),
ranging from $1.9\pm .1\ $mm to $6.1\pm .1\ $mm.
We attribute this range of values to the variability in initial
conditions imposed by loading the chamber.
%

To confirm that this front forms at the oxic-anoxic interface, we use
the oxygen-sensitive fluorescent dye
Tris(4,7-diphenyl-1,10-phenanthroline) ruthenium(II) Dichloride
($99\%$ pure, American Elements RU-OM-02).
The fluorescence of this dye is reversibly quenched by oxygen,
allowing one to visualize the distribution of
oxygen~\cite{glud1998oxygen}.
The dye is fixed within a porous plastic matrix of polyethylene
terephthalate on a $4\ $cm $\times2\ $cm Mylar sheet, as described in
ref.~\cite{larsen2011simple}. To improve the signal strength, the
detector is coated with $0.08\ $mm layer of PDMS containing $2$\%
TiO$_2$, similar to ref~\cite{klimant1995oxygen}.
This sheet is placed flush against the chamber wall. The top of the
detector is even with the sediment surface. During a measurement, the
side of the chamber is illuminated with a second high powered LED
($455\ $nm Thorlabs M455L3, not shown in fig~1) for $1$ sec. Photographing the
illuminated sheet through a $530\ $nm high pass filter allows one to
visualize the gradient in dye fluorescence and thus the oxygen
gradient.
Despite the TiO$_2$ insulation, scattering of light off of sediment
behind the detector introduces small variations in the measured oxygen
concentration
Measuring the oxygen gradient with a Clark type
microelectrode~\cite{revsbech1986microelectrodes} (Unisense OX-100)
provides independent confirmation of the gradients.

Indeed, we observe the front forms (fig 1c) near the oxic-anoxic
interface at an oxygen concentration of $c=6.7\%\pm2$\% atmospheric
concentration.
Given a diffusion coefficient of oxygen $D=4\times 10^{-6}$ cm$^2$,
the observed slope corresponds to an oxygen flux $j_o=25\pm1\ \mu$m
$\mu$M/sec.
%
After 40 hours, the increase in microbial activity reduces the
penetration depth to only $1$ mm.
The gradient then remains constant.
The scales of these gradients are typical of natural
sediments~\cite{revsbech1983microelectrode}.


The front, which extends the width of the chamber, moves towards the
surface with a constant velocity of $U=0.056\pm 0.01\ \mu$m
sec$^{-1}$.
Remarkably, the entire front moves with the same speed and does not
develop fingering instabilities~\cite{pelce2004new}..
During the motion towards the surface, we observe transient
fluctuations in the shape of the front.
The typical amplitude of these fluctuation is $\sim 300\ \mu$m,
equivalent to $\sim 3$ front widths.

The front comes to rest at a depth of $d_0=1.27\pm 0.33$ mm regardless
of the depth at which it formed.
The trajectories of fronts in six different experiments are shown in
figure~2.

Notice that front in one experimental run overshot its steady state
depth despite moving without inertia
(Reynolds number Re$=7.2\times 10^{-6}\pm3.5\times 10^{-6}$).
In this overshot experiment, we observed the transient formation of a
second, much fainter front parallel to the first (figs.~2 and 3b).
Notably, the second front appeared $0.180\pm 0.09\ $mm above the
stable front.
It persisted for $7\pm .25\ $hours before vanishing.
%
%
The return of the original front to its steady-state depth was
coincident with the disappearance of the transient front.
%
%
The presence of these double fronts in field observations has
previously been attributed to two populations of sulfur-oxidizing
microbes, a fast population that moves with the changing gradients and
a slow population that is left behind~\cite{garcia1994diel}.
As the transient front appeared discontinuously above the moving
front, rather than splitting from it, this explanation does not
explain our result.

The flow of fresh media over the sediment surface is critical to the
stability of the front.
When the flow stops, the front moves from the sediment into the
water and becomes unstable, immediately developing bioconvective
plumes (figure~\ref{front_eg}c)
\cite{platt1961bioconvection}.
Such plumes form when the density of the upward swimming microbes in the
front exceeds a critical value to develop gravity-driven
flows\cite{hillesdon1996bioconvection,hill2005bioconvection}.
Viscous forces stabilize the front in the
sediment\cite{kuznetsov20022d,kuznetsov2001numerical}.

\begin{figure}
\centering
\includegraphics[width=.43\textwidth]{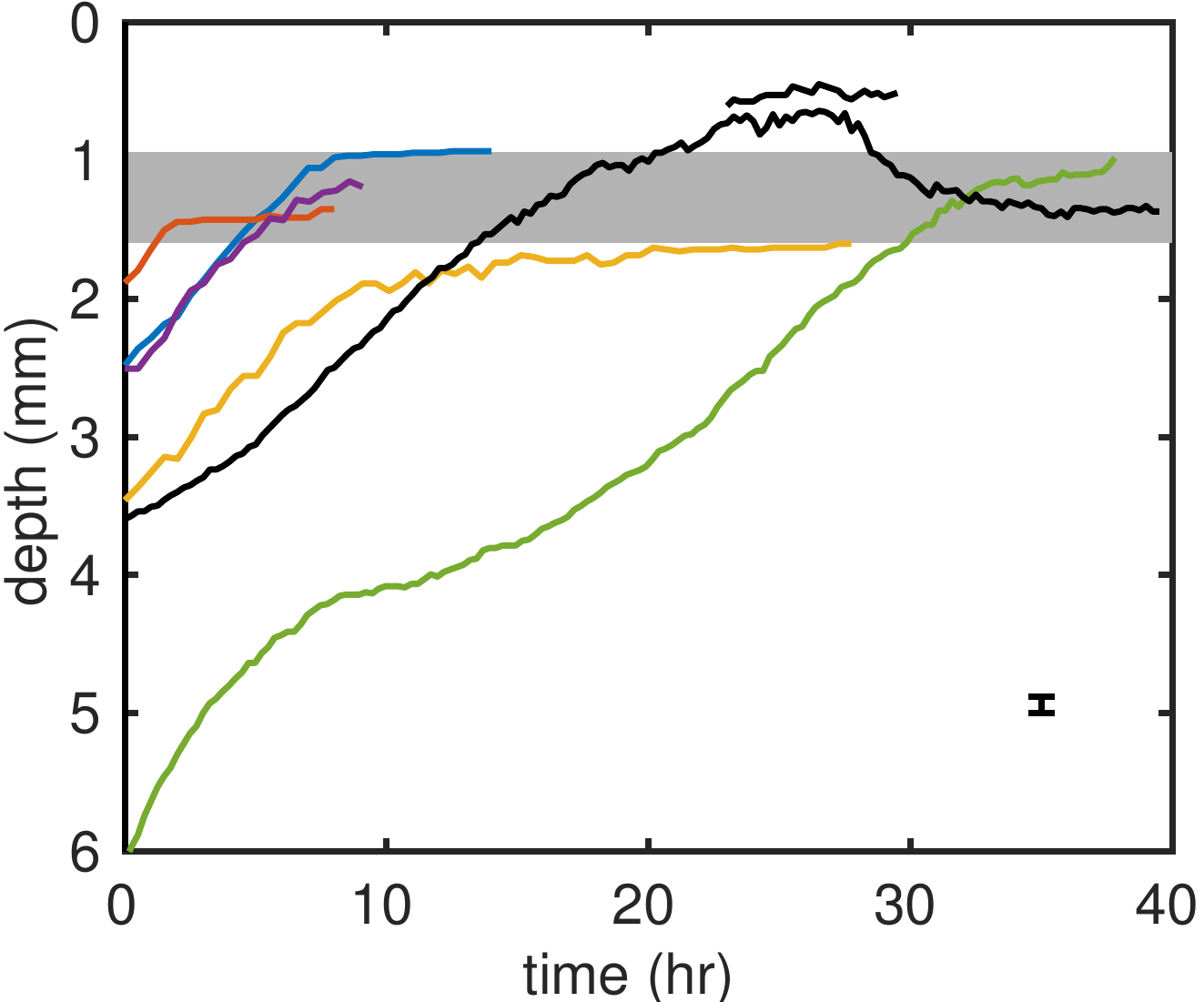}
\caption{In six experiments, fronts formed at variable depths, $2$ to
  $6$ mm below the surface. Each propagated at a constant velocity of
  $U=0.056\pm 0.01\ \mu$m sec$^{-1}$ and came to rest at a depth
  $d_0=1.27\pm 0.33$ mm (gray shaded region). The variability in $d_0$
  is similar in magnitude to the front width $0.120\pm0.060\ $mm. One
  front, marked in black, overshot its steady state depth. Near the
  surface, a second fainter front formed above it for 7 hours. Scale
  bar shows the mean width of the front.
%
}\label{exp}
\end{figure}

It is surprising that a front can form and move coherently through
this complex medium.
Past work has shown that populations of identical bacteria form fronts in
response to nutrient gradients~\cite{douarche2009coli}.
To account for this behavior, we consider how gradients of oxygen $c$
and H$_2$S (sulfide) $s$ change in response to microbial metabolism
and the resulting motion of the sulfur-oxidizing community.

We begin by considering the time $t$ evolution of these gradients
around an arbitrary distribution of microbes.
First-order kinetics for metabolism with oxygen and sulfide require 
\begin{equation}
\dt{c}=D \nabla^2 c - \sigma k(\mathbf{x},t) s c
\end{equation}
and
\begin{equation}
\dt{s}=D_s \nabla^2 s - k(\mathbf{x},t) s c.
\end{equation}
We take the diffusion coefficients of oxygen and sulfide in the
sediment (compressed in the chamber) to be $D=4\times 10^{-6}$
cm$^2$/sec and $D_s=2\times 10^{-6}$ cm$^2$/sec, respectively,
slightly smaller than their values in loose sand.
$k$ is the metabolic rate per microbe per oxygen molecule. It varies
in space with the local density of bacteria.
The stoichiometry coefficient $\sigma$ is the number of oxygen
molecules required to oxidize a sulfide \cite{kelly1999thermodynamic}.

To close these equations, one must include a model of aerotaxis to
describe how the local metabolic rate constant $k(\mathbf{x},t)$
change with the moving chemical gradients.
Two approximations make this possible for an arbitrary community of
sulfur-oxidizing microbes.
First---because the front velocity $U\approx 0.06\ \mu$m$/$sec is much
slower than the range of speeds $v\sim 1-600\ \mu$m/sec of swimming
and gliding
microbes~\cite{dworkin2006prokaryotes,purcell1977life,garcia1989rapid}---
the density of microbes and thus the metabolic rate $k$ evolves
quasistatically.
This separation of scales allows us to characterize the density and
metabolic rate of microbes in the front with their steady-state
values.
Second, we assume that the steady-state distribution of microbes is
only determined by the concentration of oxygen.
This approximation is justified as sulfur-oxidizing microbes (and the
associated protists) typically concentrate themselves near oxic-anoxic
interface, at oxygen concentrations ranging from $0$ to $10\%$
atmospheric~\cite{fenchel1996behavioural,jorgensen1983colorless,garcia1994diel}.
We describe this diversity of steady-state positions with a
distribution $p(c)$, which is the fraction of microbes that
concentrate at an oxygen concentration of $c$.
To mimic the natural range of tolerable oxygen concentrations in this
model, we take $p(c)$ to have mean $c_0=5\%$ atmospheric and variance
$\sim c_0^2$.
Combining these approximations, $k(\mathbf{x},t)=k_0
p(c(\mathbf{x},t))$, where $k_0$ is the effective metabolic rate per
oxygen molecule.
This quasistatic approximation removes any explicit dependence of the
front dynamics on the motion of bacteria in the gradients.

We observed that the front width, velocity, and steady-state depth are
similar across all experiments despite differences in their initial
conditions.
We seek to understand how this surprising consistency arises from the
length scales and timescales imposed by diffusion and metabolism.
We begin by presenting scaling relations to estimate the width,
velocity, and final front depth. We find these estimates to be roughly
consistent with observations. We then solve equations (1) and (2)
numerically to investigate the sensitivity of these dynamics to
$p(c)$.

\begin{figure*}
\centering \includegraphics[width=1\textwidth]{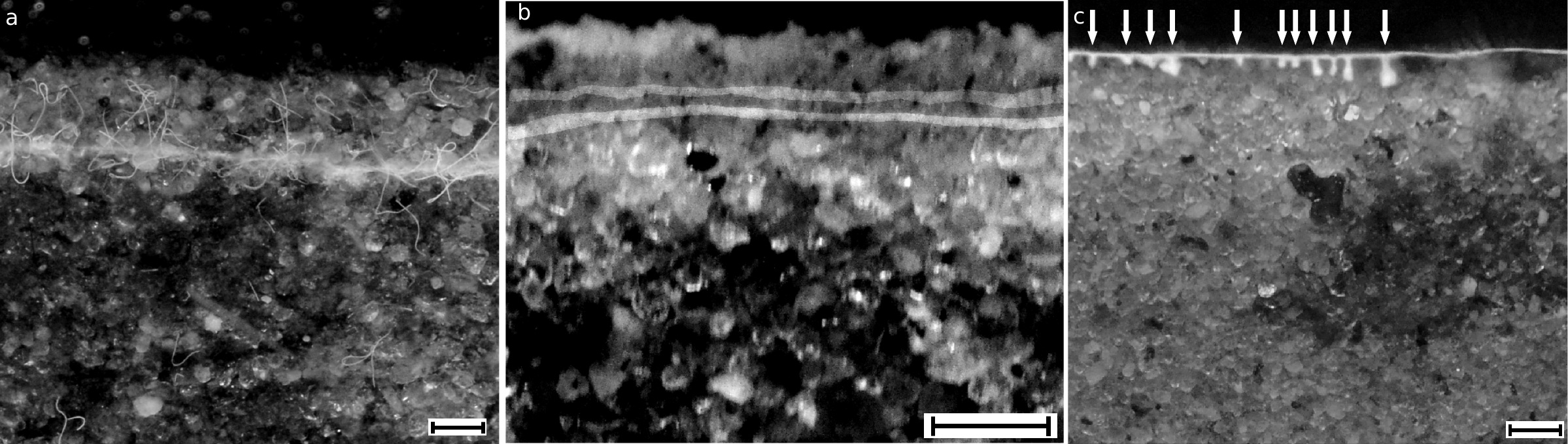}
\caption{Fronts of sulfur-oxidizing bacteria. (a) Front dynamics are
  robust to changes in bacterial community. Sediment collected in
  winter produced fronts with a high concentration of
  \textit{Beggiatoa spp.}. These filamentous bacteria are longer than the
  front thickness. (b) Parallel fronts form transiently near the
  surface. The lower front is stable, persisting for $40$ hours. The
  fainter front above it appeared as the stable front approached the
  surface and persisted for only $7$ hours before vanishing. (c) When
  the flow is stopped, the front moves to the surface. As it moves out
  of the pore space, it develops bioconvective plumes (arrows). Scale
  bars are 1 mm.}\label{front_eg}
\end{figure*}

We begin with the front width. How is it that such a thin front can
form despite being composed of diverse microbes?
The quasistatic approximation allows one to understand how diverse
microbes are forced together into a thin front by competition for
oxygen (figure~4a inset).
Metabolism sharpens oxygen gradients, leading to shorter distance
between the highest and lowest tolerable oxygen concentrations;
consequently the front thins to a width similar to the diffusive
length scale $\ell=\sqrt{D/K}$, where $K=\sigma k_0 s$ is the rate at
which oxygen is consumed in the front.
To estimate $K$, we balance the flux of oxygen
$j_o=25\pm0.1\ \mu$m$\mu$M/sec, measured from the oxygen profile in
figure 1c, with metabolic flux $K w c_0$. We find
$K=0.011\pm0.006\ $sec$^{-1}$.
The corresponding diffusive length $\ell=190\pm55\ \mu$m is indeed
similar to the measured front width $w=120\pm 60\ \mu$m.

Next, we estimate the front velocity from the aerotactic motion of
microbes in the front.
%
%
The front moves as a result of bacterial metabolism (fig.~4a inset),
which depletes oxygen near the front, and the resulting quasistatic
reorganization of bacteria in the gradient. As oxygen levels around
the front fall, the bacterial front moves to remain concentrated at
$c=c_0$.
Given a constant rate of oxygen consumption, the front moves with a
constant velocity $U$ towards the surface, where the concentration is
$c^*=170\ \mu$M (measured from the profile in figure 1c).
This propagation is analogous to the front dynamics described by the
Fisher-Kolmogorov equation, in which microbes consume a diffusing
resource at a constant rate~\cite{pelce2004new}.
However, unlike the microbial fronts described by this equation, the
metabolic rate of the sulfur-oxidizing bacteria requires the presence
of two nutrients, oxygen and sulfide, in opposing gradients.
Balancing the oxygen consumed by the moving front as $U c^*$ with the
metabolic flux $K c_0 w$, we expect a velocity of $\sim (c_0/c^*) K
w=0.240\pm0.15\ \mu$m/sec.
This value is a factor of $4$ larger than the observed velocity
$U=0.056\pm0.01\ \mu$m/s.

Finally, we consider the equilibrium position of the sulfur-oxidizing
bacteria.
As the front approaches the surface, the increasing flux of oxygen
causes it to slow.
The front reaches a steady-state depth $d_0$ where the fluxes of
oxygen $j_c\approx D (c^*-c_0)/d_0$ and sulfide $j_s$ to the bacteria
are both balanced by the metabolism.
Requiring the ratio of these fluxes to be the stochiometric ratio
$\sigma$ and solving for $d_0$, the steady-state depth is $d_0\sim D
(c^*-c_0)/\sigma j_s$.
We measure the sulfide flux from the concentration of sulfide
$s=300\pm 30\ \mu$M measured below the front (Lamotte colorimetric
sulfide assay 4456-01). This concentration corresponds to a flux of
$j_s= 25\pm 2 \mu$M $\mu$m/sec. Taking $\sigma=2$ (oxidation of H$_2$S
to SO$_4^{2-}$), we expect the front to come to rest at a depth
$2.0\pm.2\ $mm. This estimate is roughly consistent with the observed
depth $1.27\pm 0.33\ $mm.
This closes our scaling analysis.

\begin{figure}
\centering \includegraphics[width=.4\textwidth]{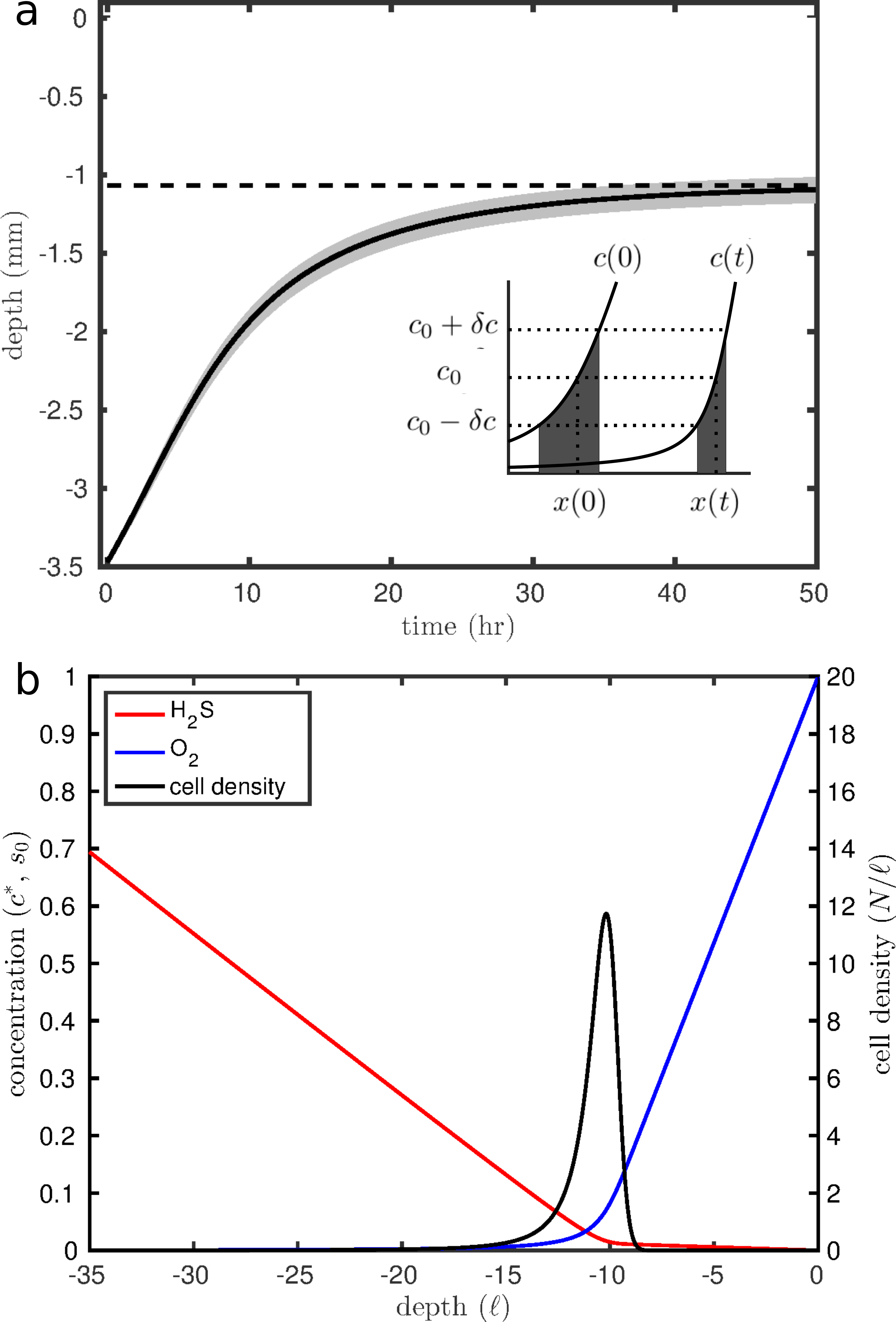}
\caption{Numerically integrating equations (1) and (2) produces a
  front that moves with an initially constant velocity to a stable
  depth $d_0$. (a) Dark line shows the trajectory of a front. The
  shaded gray region shows the predicted width of the front. The
  steady-state depth $d_0\approx 1\ $mm (dashed line) is consistent
  with observation. (inset) Sulfur-oxidizing bacteria concentrate
  within a range of low, but finite, oxygen concentrations $c=c_0\pm
  \delta c$. As metabolism sharpens oxygen gradients, the width of the
  band narrows to the diffusive length scale. As microbes consume
  oxygen, the front moves towards the surface. (b) Numerically
  integrating equations (1) and (2) generates a moving front. At the
  surface ($x=0$) $c=c^*$ and $s=0$. Sulfide is plotted relative to
  $s_0= j_s L/D_s$, the concentration in the absence of bacteria. On
  the boundary $x=-L$, oxygen vanishes ($c=0$) and sulfur-reducing
  bacteria produce a constant flux $j$ of sulfide.}\label{cartoon}
\end{figure}

To evaluate the sensitivity of these dynamics to the composition of
the microbial community, characterized by the distribution $p(c)$ of
preferred oxygen levels, we integrate equations (1) and (2)
numerically.
We initially take $p(c)$ to be a Maxwell-Boltzmann distribution with
mean $c_0= 5\%$ atmospheric.
We take the unknown parameters from the results of the scaling
analysis with no fitted parameters.
Comparing this solution (fig~4a) to the observed motion of fronts
(fig~2), shows that this solution captures the motion of the front to
the surface, its timescales, and length scales.
Figure 4b shows the distribution of oxygen, sulfide, and bacteria at
steady state.
The front dynamics depend very weakly on the distribution $p(c)$ of
steady-state oxygen concentrations of the bacteria. The trajectories
are similar if one takes $p(c)$ to be normal, exponential, or
uniformly distributed. They are similarly unchanged if one imposes
conservation of cell number.
%
The front width $w\sim\ell$ depends very weakly on the range of
tolerable oxygen concentrations. Varying $c_0$ by a factor of
$100$---from $0.003$ atmospheric to $.3$ atmospheric---results in
fronts that vary in width by a factor of $4.5$.
Thus we conclude that these dynamics are relatively insensitive to
$p(c)$. Rather the quasistatic motion of the front is determined by
the sulfide flux and surface oxygen concentration.

In conclusion, we have observed the formation of a thin front of microbes
at the oxic-anoxic interface and its constant-velocity motion to a
steady-state depth.
The front moves as the initial reservoir of oxygen (introduced to the
sediment by loading the chamber) is consumed by the microbial
population.
The front width, velocity, and steady-state position vary little
across experiments. 
We have shown their magnitudes are fixed by the metabolic timescale
$K$, diffusive length scale $\sqrt{D/K}$, sulfide flux $j_s$, and
concentrations of oxygen at the surface $c^*$.
We conclude that these dynamics arise from the coupling between
diffusion, metabolism, and aerotaxis.
The simplicity of this phenomenon is counterintuitive given the
complexity of this system.
Future work should proceed in three directions.

First, our observations have identified the transient formation of
parallel fronts (fig 3b) that cannot currently be explained.
Our model allows us to propose a hypothesis for this phenomenon: they
arise from metabolic shifts in the constituent microbes.
Notice that the steady-state depth of a front $d_0\sim D
(c^*-c_0)/\sigma j_s$ depends explicitly on stochiometric coefficient
$\sigma$, the ratio of oxygen molecules consumed per sulfide molecule.
The value of this coefficient is quantized by the oxidation state of
the waste product.  It takes values of $1/2, 3/2,$ or $2$ for
oxidation to S$^0$, SO$_3^{2-}$, and SO$_4^{2-}$, respectively.
A front of bacteria that completely oxidize sulfur ($\sigma=2$) comes to
rest at a shallower depth than those that partially oxidize sulfur
(e.g., to S$^0$).
Thus, if a sub-population in the front switch from partial to complete
sulfur oxidation (or between intermediates), they would move
discontinuously to a shallower depth.
These two fronts would then compete for sulfur and oxygen.
This hypothesis can be tested by extracting bacteria from each of the
fronts and measuring the concentration of mRNA coding for the enzymes
responsible for the different steps of sulfur oxidation.
We leave this analysis for future work.

Second, to connect these results to natural environments, we must
account for environmental fluctuations.
Our results identifies a fundamental timescale $\tau=d_0/U=6.3\pm
1.6\ $hr for microbial community to come to steady state after a
perturbation.
Comparing this timescale to the frequency $f$ of environmental
perturbations (e.g., by tides and storms), we find an important
dimensionless number $f\tau$.
If $f\tau\sim1$, as in salt marshes regularly disturbed by tides
$f=1/12$ hr$^{-1}$, the sediment never reaches steady state. We
therefore expect that the decay of organics is strongly coupled to
environmental fluctuations.
%
To measure the relationship between organic decay and environmental
fluctuations, we will modify this experiment (fig~1a) to include a
variable flow of fresh media over the surface and measure resulting
variability in gross metabolic rate.

Finally, we have only examined the dynamics of the oxic-anoxic
interface in a fixed quantity of sediment.
Including a sedimentation rate of $U_s$ at the surface, the
sediment-water interface moves away from the microbial front.
To examine the influence of this parameter, we will modify our
experiment to include a flux of sediment and measure the gross
metabolic rate.
These dynamics may explain the observed variability of cell density
with sedimentation rate in the sea~\cite{kallmeyer2012global}.

This work was supported by HFSP RGP0037. The authors declare that they have no
competing financial interests. Correspondence and requests for materials
should be addressed to A.P.P~(email: apetroff@rockefeller.edu).


%

\end{document}